\documentclass[prl,twocolumn,aps,amsmath,superscriptaddress,nofootinbib]{revtex4}
\usepackage{graphicx}
\usepackage{bm}
\usepackage{epsfig}

\newif\ifpdf
\ifx\pdfoutput\undefined
\pdffalse 
\else
\pdfoutput=1 
\pdftrue
\fi

\def\bsigma{\mbox{\boldmath $\sigma$}}

\def\OMIT#1{}

\newcommand{\nn}{\nonumber} 

\newcommand{\bn}{{\bar n}}
\newcommand{\bea}{\begin{eqnarray}}
\newcommand{\eea}{\end{eqnarray}}

\newcommand{\bnP}{\bar {\cal P}}
\newcommand{\ppP}{{\cal P}_\perp}

\newcommand{\cP}{{\cal P}}

\newcommand{\mups}{M_\Upsilon}

\begin{document}

\ifpdf
\DeclareGraphicsExtensions{.pdf, .jpg}
\else
\DeclareGraphicsExtensions{.eps, .jpg,.ps}
\fi


\preprint{\vbox{ \hbox{CMU-HEP02-14}   \hbox{FERMILAB-Pub-02/296-T} }}

\title{The Resummed Photon Spectrum in Radiative $\Upsilon$ Decays} 

\author{Sean Fleming}
\affiliation{Department of Physics, Carnegie Mellon University,
      	Pittsburgh, PA 15213
	\vspace{0.1cm}}
	
\author{Adam K. Leibovich}
\affiliation{Theory Group, Fermilab, 
	P.O. Box 500, 
	Batavia, IL 
	60510
	}




\begin{abstract}
We present a theoretical prediction for the photon spectrum in
radiative $\Upsilon$ decay including the effects of resumming the
endpoint region, $E_\gamma \to \mups/2$. Our approach is based on
NRQCD and the soft collinear effective theory. We find that our
results give much better agreement with data than the leading order
NRQCD prediction.
\end{abstract}

\maketitle

\newpage

The radiative decay $\Upsilon\to X\gamma$ was first investigated about
a quarter century ago \cite{firstRad}.  The conventional wisdom at
was that this process is computable in perturbative QCD due
to the large mass of the $b$ quarks.  Since then, we have learned much
about quarkonium in gerneral \cite{bbl} and this process in particular
\cite{Catani:1995iz,Maltoni:1999nh,Rothstein:1997ac,
Kramer:1999bf,Bauer:2001rh}.  In addition, CLEO is currently taking
data on the low lying $\Upsilon$ resonances and will soon be able to 
update their original measurement of this decay \cite{Nemati:1996xy}.
It is thus timely to reexamine the theoretical predictions for this
rate.

The current method for calculating the direct radiative decay of the
$\Upsilon$ is by using the operator product expansion (OPE), with the
operators scaling as some power of the relative velocity of the heavy
quarks, $v$, given by the power counting of Non-Relativistic QCD
(NRQCD) \cite{bbl}.  The $v\to0$ limit of NRQCD coincides with the
color-singlet (CS) model calculation of~\cite{firstRad}.

This picture is only valid in the intermediate range of the photon
energy ($0.3 \lesssim z \lesssim 0.7$, where $z = 2 E_\gamma
/M$, and $M = 2m_b$).  In the lower range, $z\lesssim 0.3$,
photon-fragmentation contributions are important~\cite{Catani:1995iz,
Maltoni:1999nh}. At large values of the photon energy, $ z\gtrsim
0.7$, both the perturbative expansion~\cite{Maltoni:1999nh} and the
OPE~\cite{Rothstein:1997ac} break down.

The breakdown at large $z$ is due to NRQCD not including collinear
degrees of freedom.  The correct effective field theory is a
combination of NRQCD for the heavy degrees of freedom and the
soft-collinear effective theory
(SCET)~\cite{Bauer:2001ew,Bauer:2001ct} for the light
degrees of freedom.  In a previous paper~\cite{Bauer:2001rh} we
applied SCET to the color-octet (CO) contributions to radiative
$\Upsilon$ decay.  Here we treat the CS contribution at the endpoint
within SCET.  In this letter we present the main results of the
analysis, and leave the details to a companion
paper~\cite{FL}.

The inclusive photon spectrum can be written as a sum of a direct and
a fragmentation contribution~\cite{Catani:1995iz},
\begin{equation}
\frac{d\Gamma}{dz}= \frac{d\Gamma^{\rm dir}}{dz}
  +\frac{d\Gamma^{\rm frag}}{dz},
\end{equation}
where in the direct term the photon is produced in the hard
scattering, and in the fragmentation term the photon fragments from a
parton produced in the initial hard scattering.  The fragmentation
contribution has been well studied in Ref.~\cite{Maltoni:1999nh}, and
we do not add anything new here.  

The direct contribution can be calculated using the OPE, where the rate
can be written as
\begin{equation} \label{nrqcdope}
\frac{d \Gamma}{d z} = \sum_n C_n (M,z) 
   \langle \Upsilon \vert {\cal O}_n \vert \Upsilon \rangle \,.
\end{equation}
The $C_i$ are short-distance coefficients, calculable as a 
perturbative series in $\alpha_s(M)$, and the ${\cal O}$ are NRQCD
operators, scaling with specific powers of $v\ll1$.

At leading order in $v$ only a CS term contributes. The CS operator
${\cal O}(1,{}^3S_1)$ creates and annihilates a CS quark-antiquark
pair in a ${}^3S_1$ configuration, and is multiplied by the CS
coefficient, which at leading order is proportional to
$\alpha^2_s(M)$.  There are also CO contributions down by  $v^4$. Two
of these, proportional to the CO $^1S_0$ and $^3P_0$ matrix elements
(MEs), give rise to large enhancement at the endpoint
\cite{Maltoni:1999nh}.  Since the CO $^1S_0$ and $^3P_0$ MEs are
unknown, and since the data does not show any enhancement near the
upper endpoint, we set them to zero.  However, we include the CO
$^3S_1$ ME.  It has a sizable fragmentation contribution, but becomes
negligible as $z$ increases, and thus does not conflict with
data~\cite{Maltoni:1999nh}.

The lowest order CS direct rate is~\cite{firstRad}
\begin{eqnarray}\label{LOrate}
\frac1{\Gamma_0} \frac{d\Gamma^{\rm dir}_{\rm LO}}{dz} &=&  
\frac{2-z}{z} + \frac{z(1-z)}{(2-z)^2} +
2\frac{1-z}{z^2}\ln(1-z) \nn\\
&& - 2\frac{(1-z)^2}{(2-z)^3} \ln(1-z),
\end{eqnarray}
where 
\begin{equation}
\Gamma_0 = \frac{32}{27}\alpha\alpha_s^2e_b^2
\frac{\langle\Upsilon\vert{\cal O}_1(^3S_1)\vert\Upsilon\rangle}{m_b^2},
\label{gamma0}
\end{equation}
and $e_b = -1/3$.  The ME is related to the wavefunction
\begin{equation}\label{singletWF}
\langle \Upsilon \vert {\cal O}_1(^3S_1) \vert \Upsilon \rangle =
  \frac{N_c}{2\pi} |R(0)|^2.
\end{equation}
The $\alpha_s$ correction to this rate was calculated numerically in
Ref.~\cite{Kramer:1999bf}, leading to small corrections over most of
phase space. In the endpoint region, however, the corrections are of
order the leading contribution.

In the endpoint region, the outgoing gluons are moving back-to-back to
the photon, with large energy and small invariant mass (ie, a
collinear jet).  We must, therefore, couple NRQCD to
SCET~\cite{Bauer:2001ct}.  The scales, set by the
lightcone momentum components of the collinear particles, are widely
separated.  If we choose $p^-$ to be ${\cal O}(M)$, then $p_\perp/p^-
\sim \lambda$, and $p^+/p^- \sim \lambda^2$, where $\lambda$ is a
small parameter.  Here the collinear scale is
\begin{eqnarray}
\mu_c \sim M\sqrt{1-z}\sim \sqrt{M \Lambda_{\rm QCD}}\,.
\end{eqnarray}
Thus $\lambda$ is of order $\sqrt{1-z}\sim\sqrt{\Lambda_{\rm QCD}/M}$.

There are two types of fundamental objects in SCET (fields and Wilson
lines) and two separate sectors (collinear and usoft).  In the
collinear sector there is a fermion field $\xi_{n,p}$, a gluon field
$A_{n,q}^\mu$, and a Wilson line
\begin{equation}
W_n(x)=
 \bigg[ \sum_{\rm perms} {\rm exp} 
  \left( -g_s \frac{1}{\bnP} \bn \cdot A_{n,q}(x) \right) \bigg] \,.
\end{equation}
Collinear fields are labeled by a direction $n^\mu$ and 
the large components  ($\bn\cdot q, q_\perp$). 
The operator ${\cal P}^\mu$ projects out the momentum label. 
Likewise in the usoft sector there is a
fermion field $q_{us}$, a gluon field $A^\mu_{us}$, and a Wilson line
$Y$. Operators are constructed out of these objects such that they are
gauge invariant.  Thus, operators with collinear gluons are built out
of the homogeneous (order $\lambda$) component of the collinear
field strength, $\bnP B^\mu_\perp\equiv\bn_\nu G^{\nu\mu}_n$
\cite{Bauer:2002nz},
\begin{equation}\label{bfield}
B^\mu_\perp =  \frac{-i}{g_s} W^\dagger (\ppP^\mu + g_s (A^\mu_{n,q})_\perp)W. 
\end{equation}

We now write down the leading operator.  Aside from $B_\perp$, we
also need the NRQCD heavy quark and antiquark fields, $\psi_{\bf
p}$ and $\chi_{-{\bf p}}$, which transform only under usoft (not
collinear) gauge transformations.  A CS ${}^3S_1$ $b\bar{b}$ pair
decays into a photon and a colorless jet of gluons.  We must,
therefore, include two of the $B_\perp$ fields in a colorless
configuration, and the only operator is
\begin{eqnarray}\label{3s1op1}
{\cal O}(1,{}^3S_1) &=&
\\
& &  \hspace{-5ex}
 \chi^\dagger_{-{\bf p}} \bsigma^\delta \psi_{\bf p}
{\rm Tr} \big\{ B^\alpha_\perp \, 
\Gamma^{(1,{}^3S_1)}_{\alpha \beta \delta \mu} ( \bnP, \bnP^\dagger ) \, 
B^\beta_\perp \big\} \,,
\nn
\end{eqnarray}
where $\bnP^\dagger$ acts to the left.  
Momentum conservation forces the momentum of the
jet to be $M$, so $B^\alpha_\perp ( \bnP+ \bnP^\dagger)
B^\beta_\perp = -M B^\alpha_\perp B^\beta_\perp$.  Introducing
$\cP_{-} = \bnP- \bnP^\dagger$, Eq.~(\ref{3s1op1}) becomes
\begin{eqnarray}\label{3s1op2}
{\cal O}(1,{}^3S_1)(M)& = &
\\
& & \hspace{-7ex}
\chi^\dagger_{-{\bf p}} \bsigma^\delta \psi_{\bf p}
{\rm Tr} \big\{ B^\alpha_\perp \, 
\Gamma^{(1,{}^3S_1)}_{\alpha \beta \delta \mu} ( M,\cP_{-} ) \, 
B^\beta_\perp \big\} \,.
\nn
\end{eqnarray}
%
%
%
Matching onto QCD at tree level, we obtain
\begin{equation}\label{matching1}
\Gamma^{(1,{}^3S_1)}_{\alpha \beta \delta \mu} ( M,\bn\cdot q_{-})
= \frac{4 g_s^2 e e_b}{3 M} g_{\alpha\beta}^\perp g_{\mu\delta} \,,
\end{equation}
for a transverse photon, where $\bn\cdot q_{-} = \bn\cdot q-
\bn\cdot q'$ and $g_\perp^{\mu\nu} = g^{\mu\nu} -
(n^\mu\bar n^\nu + n^\nu\bar n^\mu)/2$.

The inclusive $\Upsilon \to X \gamma$ rate can be factored into hard,
jet, and usoft functions at the endpoint.  Using the optical theorem
the inclusive spectrum can be written as
\begin{equation}\label{optcthm}
\frac{d \Gamma}{d z} = z \frac{M}{16 \pi^2} {\rm Im} T(z)  \,,
\end{equation}
where the forward scattering amplitude $T(z)$ is 
\begin{equation}\label{fsamp}
 T(z) = -i \int d^4 x e^{-iq\cdot x} 
\langle \Upsilon | T J^\dagger_\mu (x) J_\nu(0) | \Upsilon \rangle 
   g^{\mu \nu}_\perp \,.
\end{equation}
The $T$ indicates time ordering. Matching onto SCET the forward
scattering amplitude can be written as
\begin{equation}\label{morebots}
T(z) = \sum_\omega H(\omega,\mu) T_{\rm eff} (\omega,z,\mu) \,,
\end{equation}
where
\begin{equation}
T_{\rm eff} (\omega,z,\mu) =
\int  d \ell^+  J_\omega[\ell^+ + M(1-z)] S(\ell^+) \,.
\end{equation}
After decoupling usoft degrees of freedom \cite{Bauer:2001ct}, the CS
jet function is defined as
\begin{eqnarray}\label{cscoll1}
 \langle 0 | T \, {\rm Tr}\big[ B^{(0) \alpha}_\perp 
  \delta_{\omega,\cP_-} B^{(0) \beta}_\perp \big] (x) && \hspace{-3ex}
{\rm Tr}\big[ B^{(0) \alpha'}_\perp \delta_{\omega',\cP_-} 
  B^{(0) \beta'}_\perp \big] (0) |0 \rangle 
\nn \\
& & \hspace{-30ex} \equiv
 \frac{i}{2} 
(g^{\alpha \alpha'}_\perp g^{\beta \beta'}_\perp+ 
  g^{\alpha \beta'}_\perp g^{\beta \alpha'}_\perp)
\, \delta_{\omega, \omega'} \, \int \frac{d^4 k}{(2 \pi)^4} 
e^{-i k\cdot x} J_\omega(k^+) \,,
\nn \\
\end{eqnarray}
and the CS usoft function is defined as
\begin{eqnarray}\label{cssoftfn}
S(\ell^+) &=& \int \frac{d x^-}{4 \pi} e^{\frac{-i}{2} \ell^+ x^-}
\\ 
& &  \times
 \langle \Upsilon | 
T \big[\psi^\dagger_{\bf p} \bsigma_i \chi_{-{\bf p}} \big] (x^-)
\big[\chi^\dagger_{-{\bf p}'} \bsigma_i \psi_{{\bf p}'} \big] (0) | 
\Upsilon \rangle \, \nn\\
&&\label{soft}
\hspace{-5ex}= \langle \Upsilon | \psi^\dagger_{\bf p} \bsigma_i \chi_{-{\bf p}}
\delta(i n\cdot \partial - \ell^+) 
\chi^\dagger_{-{\bf p}'} \bsigma_i \psi_{{\bf p}'} | \Upsilon \rangle \,.
\end{eqnarray}
The hard coefficient $H(\omega,\mu)$ can be calculated perturbatively
in an expansion in $\alpha_s(M)$. At tree level we obtain
\begin{equation}
H(\omega,\mu) = \frac{4}{3}\left(\frac{4g_s^2 e e_b}{3M}\right)^2\,.
\end{equation}

\begin{figure}[t]
\centerline{ \includegraphics[width=1.5in]{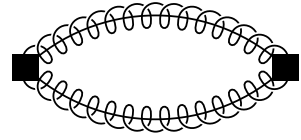}}
\caption{\it Feynman diagram for the leading order jet function.
Collinear gluons are represented by a spring with a line.
\label{vacloop}}
\end{figure}

At the collinear scale $\mu_c$ we perform an OPE, integrate out
collinear modes and match onto a non-local usoft operator,
Eq.~(\ref{cssoftfn}), convoluted with a Wilson coefficient,
\begin{equation}\label{nonotmore}
T(z) = \int d\ell^+ S(\ell^+) {\cal H}_J[\ell^+ + M(1-z)] \,.
\end{equation}
To leading order in $\alpha_s(M\sqrt{1-z})$, the jet function is calculated
from the Feynman diagram shown in Fig.~\ref{vacloop}.
Evaluating the diagram gives 
\begin{eqnarray}
J_\omega(k^+) &=& \frac{\Gamma(\epsilon)}{8 \pi^2} 
\left(4 \pi \frac{\mu^2}{-M^2-i\delta}\right)^\epsilon 
\\
& & 
\times \int^1_{-1} d \xi \, \frac{1}{[(k^+/M)(1-\xi^2)]^\epsilon} \,  
\delta_{\omega,M \xi} \,,
\nn
\end{eqnarray}
and taking the imaginary part we obtain
\begin{equation}
{\rm Im } J_\omega(k^+) =  \frac{1}{8 \pi} \Theta(k^+)
 \int^1_{-1} d \xi \delta_{\omega,M \xi} \,.
\end{equation}
Combining, we get
\begin{eqnarray}\label{noouch}
{\rm Im} T(z) &=& \frac{2 M}{M^2} 
\int d \ell^+ \, S(\ell^+) \, \Theta[\ell^++M(1-z)]
\nn\\
 & & \times  
\, \frac{8 \pi}{3} \bigg( \frac{4 \alpha_s(M) e e_b}{3 M} \bigg)^2 
\int_{-1}^1 d\xi,
\end{eqnarray}
where the $2M/M^2$ accounts for the non-relativistic normalization of
the $\Upsilon$ state in the usoft function.  This is precisely in form
given in Eq.~(\ref{nonotmore}), and it is straightforward to read off
${\cal H}_J$.

Using Eq.~(\ref{soft}), we can integrate over $\ell^+$, giving
\begin{eqnarray}\label{nopenouch}
{\rm Im} T(z) &=&
\frac{16 \pi^2}{M} \bigg( 
\frac{32 \alpha \alpha^2_s(M) e_b^2}{27 m_b^2} \bigg)
\\   
& & \hspace{-10ex} \times  
\langle \Upsilon | \psi^\dagger_{\bf p} \bsigma_i \chi_{-{\bf p}}
\Theta[i n\cdot \partial+M(1-z)] 
\chi^\dagger_{-{\bf p}'} \bsigma_i \psi_{{\bf p}'} | \Upsilon \rangle
\nn \\
\label{nofinalfsa}
&=& \Theta(\mups - M z) 
\, \frac{16 \pi^2}{M }\Gamma_0,
\end{eqnarray}
where we used the results of Ref.~\cite{Rothstein:1997ac} for the
final line.  Plugging into Eq.~(\ref{optcthm}) gives the $z\to 1$
limit of Eq.~(\ref{LOrate}).

\begin{figure}[t]
\centerline{\includegraphics[width=1.5in]{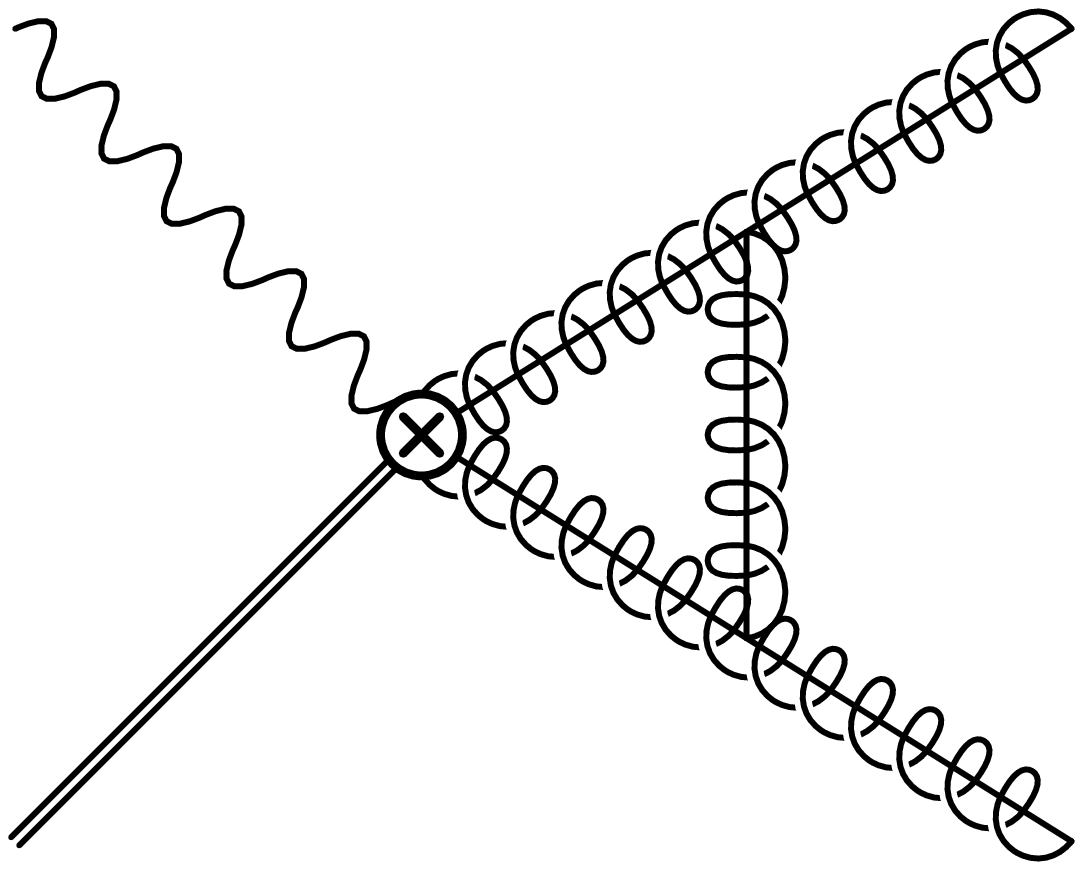}\qquad
\includegraphics[width=1.5in]{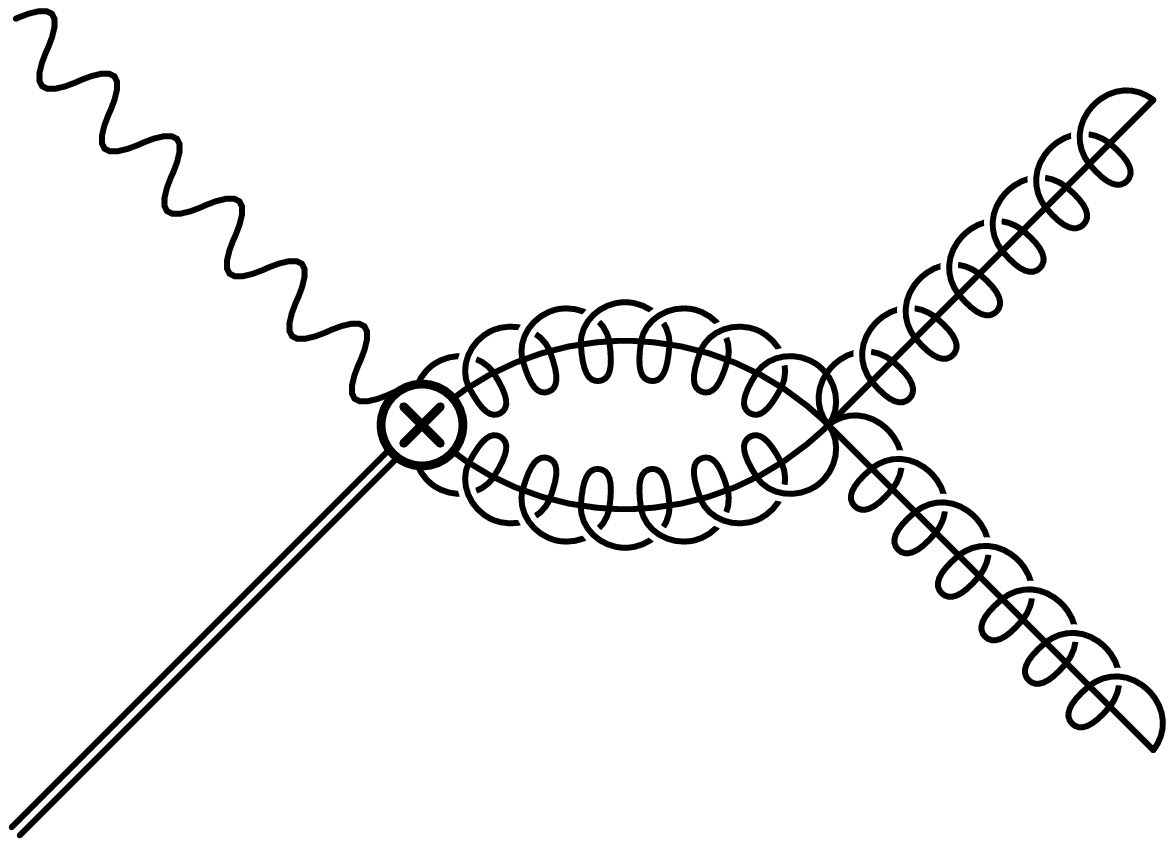}}
\centerline{\includegraphics[width=1.5in]{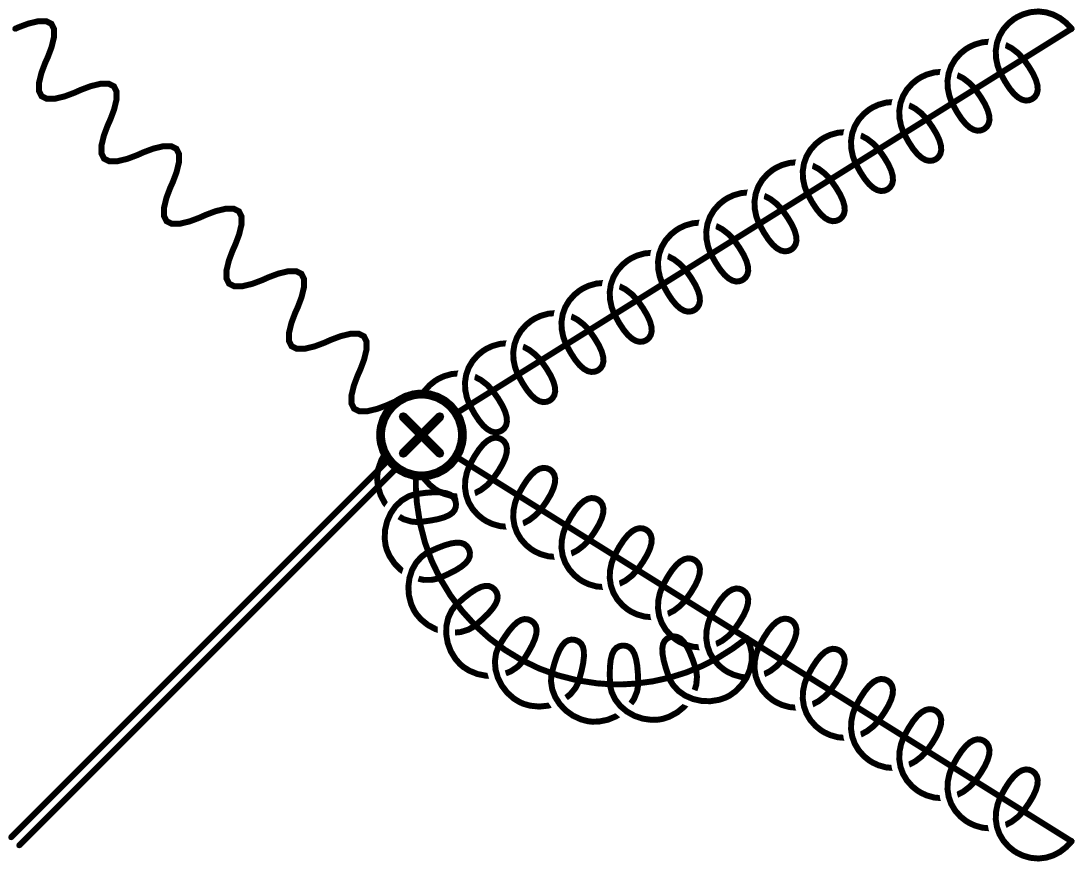}\qquad
\includegraphics[width=1.5in]{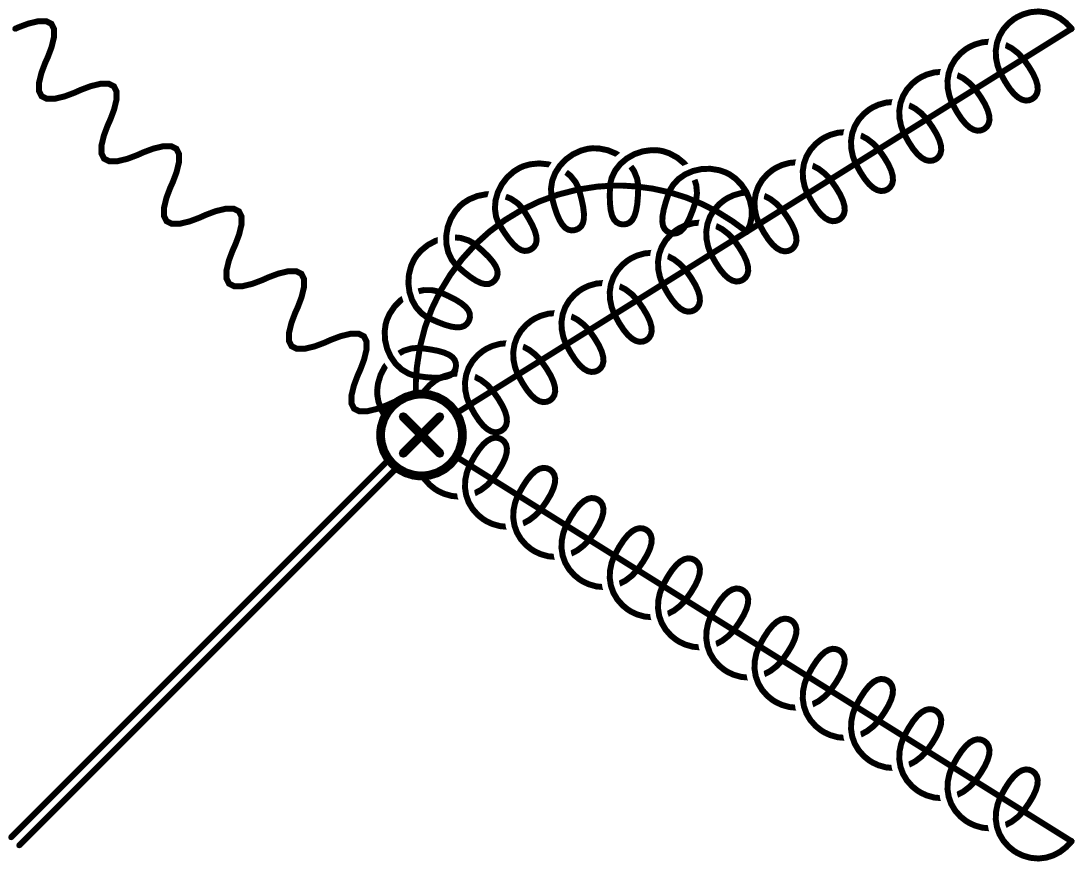}}
\caption{\it Diagrams needed to calculate the CS counterterm.
\label{oneloop}}
\end{figure}

At this point, large logarithms will appear in the jet function at
higher order. This can be avoided by running operators from $M$ to
$\mu_c$, which sums logs of $1-z$. To run the CS operator, we
calculate the counter term, determine the anomalous dimension, and use
this in the renomalization group equations (RGEs). The graphs needed
are shown in Fig.~\ref{oneloop}.
Diagrams involving usoft gluons vanish. Feynman rules for the
vertex operators are given in Ref.~\cite{FL}.  We perform our 
calculation in Feynman gauge, and obtain a relatively
simple result for the one-loop UV-divergent term  
\begin{eqnarray}\label{UVterms}
{\cal A} &=& 
\frac{1}{\epsilon} \, \sum_{\omega } {\cal O}(1,{}^3S_1)(\omega)
\frac{\alpha_s(\mu) C_A}{2 \pi} \bigg[ 1
\\
&& \hspace{-5ex} + \frac{M^2+ \omega^2 }{M^2}
 \bigg(\frac{M}{M+\omega} \ln\frac{M-\omega} {2M}
 + \frac{M}{M-\omega} \ln\frac{M+\omega}{2M}\bigg) \bigg] . 
\nn
\end{eqnarray}
This depends on the large momentum component of the gluons.  
The divergent piece must be
canceled by the counterterm $Z_3 / Z_{\cal O}-1$, where $Z_{\cal O}$
is the CS vertex counterterm, and $Z_3$ is the gluon
wavefunction counterterm
\begin{equation}
Z_3 = 1 + \frac{\alpha_s}{4 \pi} \frac{1}{\epsilon} 
\left( C_A \frac{5}{3} -n_f \frac{2}{3} \right) \,.
\end{equation}

The anomalous dimension is obtained through the standard method, and
the RGE for the coefficient is
\begin{equation}
\mu \frac{d}{d \mu} \Gamma^{(1,{}^3S_1)}(\mu, \omega) = 
\gamma(\mu, \omega) \Gamma^{(1,{}^3S_1)}(\mu, \omega) \,.
\end{equation}
Solving this equation gives
\begin{eqnarray}\label{resumcoeff}
\ln \! \bigg( \frac{\Gamma^{(1,{}^3S_1)}(\mu, \omega)}
    {\Gamma^{(1,{}^3S_1)}(M, \omega)} \bigg) &=& 
 \\
& & \hspace{-20ex}
\frac{2}{\beta_0} \bigg\{ C_A 
\bigg[ \frac{11}{6} + \frac{M^2+ \omega^2 }{M^2}
\bigg(\frac{M}{M+\omega} \ln\frac{M-\omega}{2M} 
\nn \\ 
& &   \hspace{-15ex} + \frac{M}{M-\omega} \ln\frac{M+\omega}{2M}\bigg) \bigg] 
 - \frac{n_f}{3}\bigg\}  \ln\!\bigg( \frac{\alpha_s(\mu)}{\alpha_s(M)}\bigg).
\nn
\end{eqnarray}
Logarithms of the form $\ln(\mu/M)$ have been summed into
$\Gamma^{(1,{}^3S_1)}(\mu, \omega)$, and any logarithms in the
operator are of the form $\ln(\mu_c/\mu)$.  If we take $\mu\sim\mu_c$
all large logarithms of the ratio $\mu_c/M$ will sit in the
coefficient.

We now obtain the resummed rate, by substituting
Eq.~(\ref{resumcoeff}) into Eq.~(\ref{noouch}), giving
\begin{eqnarray}\label{ouch}
{\rm Im} T(z) &=&  2 M \int d \ell^+ \, S(\ell^+) \, \Theta[\ell^++M(1-z)]
\\
& & \hspace{-10ex}  \times 
 \frac{16 \pi}{3} \bigg( \frac{4 \alpha_s(M) e e_b}{3 M^2} \bigg)^2 
\int_0^1 d \eta \bigg[ \frac{\alpha_s(M\sqrt{1-z})}{\alpha_s(M)} 
\bigg]^{2\gamma(\eta)},
\nn
\end{eqnarray}
where $\eta = 1/2(\xi + 1)$ and 
\begin{eqnarray}\label{anom}
\gamma(\eta) & \equiv &  \frac{2}{\beta_0} \bigg\{ C_A \bigg[ \frac{11}{6}
\\
& & \hspace{-7ex}
+\big(\eta^2 + (1-\eta)^2 \big) \bigg( \frac{1}{1-\eta} \ln \eta
+ \frac{1}{\eta}\ln (1-\eta) \bigg) \bigg] -\frac{n_f}{3} \bigg\}.
\nn
\end{eqnarray}
Again integrating over $\ell^+$ and inserting into
Eq.~(\ref{optcthm}), the resummed CS contribution to the
decay rate is,
\begin{equation}\label{resrate}
\frac1{\Gamma_0}\frac{d\Gamma_{\rm resum}}{dz} =
z \int_0^1 d \eta \bigg[ \frac{\alpha_s(M\sqrt{1-z})}{\alpha_s(M)}
\bigg]^{ 2\gamma(\eta) } .
\end{equation}
We can expand in $\alpha_s(M)$ to obtain an analytic
expression for the next-to-leading logarithmic contribution
\begin{equation}\label{expresrate}
\frac{1}{\Gamma_0} \frac{d \Gamma}{d z} = 
z \bigg\{ 1 + \frac{ \alpha_s}{6 \pi} \big[ C_A (2 \pi^2 - 17)+2 n_f 
\big] \ln(1-z) \bigg\} \,.
\end{equation}
As $z$ approaches one the ${\cal O}(\alpha_s)$ term becomes of order
one, precisely the behavior observed in Ref.~\cite{Kramer:1999bf}. The 
resummed result does not suffer from this problem.

We now combine the different contributions to obtain a prediction for
the photon spectrum.  We will marry our expression for the CS spectrum
in the endpoint with the leading order result by interpolating between
the two
\begin{equation}\label{fulleq}
\frac{1}{\Gamma_0} \frac{d \Gamma_{\rm int}}{d z}= 
\bigg( \frac{1}{\Gamma_0} \frac{d \Gamma_{\rm LO}^{\rm dir}}{d z} - z \bigg)
+ \frac{1}{\Gamma_0} \frac{d \Gamma_{\rm resum}}{d z} \,.
\end{equation}

Before we proceed we need the NRQCD MEs.  We can extract the CS ME
from the $\Upsilon$ leptonic width. The CO MEs are more difficult to
determine.  NRQCD predicts that the CO MEs scale as $v^4$ compared to
the CS ME.  In Ref.~\cite{Petrelli:1998ge} it was argued that an extra
factor of $1/2 N_c$ should be included.  We set the $^1S_0$ and
$^3P_0$ MEs to zero, and the $^3S_1$ ME to $\langle\Upsilon\vert{\cal
O}_8(^3S_1)\vert\Upsilon\rangle = v^4 \langle\Upsilon\vert{\cal
O}_1(^3S_1)\vert\Upsilon\rangle$, where we use $v^2 = 0.08$.

The CLEO collaboration measured the number of photons in inclusive
$\Upsilon(1S)$ radiative decays~\cite{Nemati:1996xy}.  The data does
not remove the efficiency or energy resolution and is the number of
photons in the fiducial region, $|\cos\theta|<0.7$.  In order to
compare our theoretical prediction to the data, we integrate over the
barrel region and convolute with the efficiency that was modeled in
the CLEO paper.  We do not do a bin-to-bin smearing of our prediction.
\begin{figure}[t]
\vspace{-1.4cm}
\centerline{\includegraphics[width=3.7in]{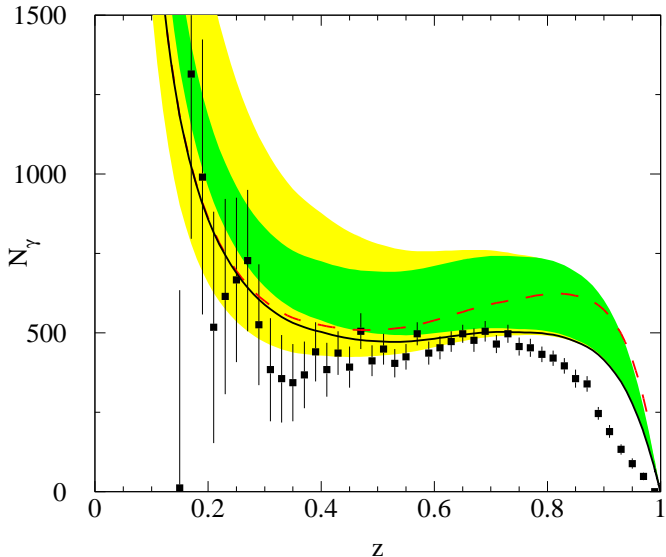}}
\caption{\it The inclusive photon spectrum, compared with data
\cite{Nemati:1996xy}.  The theory predictions are described in the
text.}
\label{comparedata}
\end{figure}

In Fig.~\ref{comparedata} we compare our prediction to the data.  The
error bars on the data are statistical only.  The dashed line is the
direct tree-level plus fragmentation result, while the solid curve
includes the resummation in Eq.~(\ref{fulleq}).  For these two
curves we use the $\alpha_s$ extracted from these data,
$\alpha_s(M_\Upsilon) = 0.163$, which corresponds to $\alpha_s(M_Z) =
0.110$ \cite{Nemati:1996xy}.  The shape of the resummed result is much
closer to the data than the tree-level curve, though it is not a perfect fit.
We also show the Eq.~(\ref{fulleq}) plus fragmentation
result, using the PDG value of $\alpha_s(M_Z)$, including theoretical
uncertainties, denoted by the shaded region.  To obtain the darker
band, we first varied the choice of $m_b$ between $4.7 {\rm\ GeV} <
m_b < 4.9 {\rm\ GeV}$ and the value of $\alpha_s$ within the errors
given in the PDG, $\alpha_s(M_Z) = 0.1172(20)$ \cite{Hagiwara:pw}.
Varying $m_b$ and $\alpha_s$ modifies the extraction of the CS ME from
$3.31 {\rm\ GeV}^3$ to $3.56 {\rm\ GeV}^3$.  We also varied the
collinear scale, $\mu_c$ from $M\sqrt{(1-z)/2} < \mu_c <
M\sqrt{2(1-z)}$.  Finally, the lighter band also includes the
variation, within the errors, of the parameters for the quark to
photon fragmentation function extracted by ALEPH
\cite{Buskulic:1995au}.  The low $z$ prediction is dominated by the
quark to photon fragmentation coming from the CO $^3S_1$ channel.  We
did not assign any error to the CO $^3S_1$ ME.  Since it is unknown,
there is a very large uncertainty in the lower part of the prediction
that we decided not to show. Note that the CO ${}^1S_0$ and ${}^3P_0$
contribution increases the theoretical prediction at the upper
endpoint~\cite{FL}.  It is thus clear the data favors a very small
value for the CO ${}^1S_0$ and ${}^3P_0$ MEs.  This is why we set
these to zero in our analysis.  Negative values for these MEs are
possible, and would give a bit better fit to the shape.

\acknowledgments 
We would like to thank C.~Bauer, D.~Besson, R.~Briere, I.~Rothstein,
and I.~Stewart for helpful discussions. This work was supported in
part by the DoE under grant numbers DOE-ER-40682-143
and DE-AC02-76CH03000.


\end{document}